\documentclass[epj,referee]{svjour}
\usepackage{graphics}

\title{Aging and response properties in the parking-lot model}
\author{J. Talbot\inst{1,2}, G. Tarjus\inst{2}, and P. Viot\inst{2,3}} 
\institute{\inst{1}
Department  of   Chemistry  and   Biochemistry,  Duquesne  University,
Pittsburgh, PA 15282-1530, USA \\ \inst{2}Laboratoire de Physique Th{\'e}orique
des Liquides, Unit{\'e} Mixte de Recherche  UMR 7600, Universit{\'e} Pierre et
Marie Curie, 4, place Jussieu 75252 Paris, Cedex, 05 France
\\\inst{3} Laboratoire de
Physique Th{\'e}orique,   Unit{\'e}  Mixte  de   Recherche UMR 8627, Bat.      210,
Universit{\'e} de Paris-Sud 91405 ORSAY Cedex France }

\begin{document} 

\abstract{
An   adsorption-desorption   (or  parking-lot)   model can   reproduce
qualitatively the densification kinetics and    other features of    a
weakly vibrated granular material. Here we study the the two-time
correlation  and response functions  of the model and demonstrate that
their behavior is consistent with recently observed memory effects in
granular materials. Although the densification kinetics and hysteresis
are robust properties,    we show that   the  aging  behavior of   the
adsorption-desorption model is different from other models of granular
compaction.    We  propose an experimental   test   to distinguish the
possible aging behaviors.
}
\PACS{05.70.Ln,81.05.Rm,75.10.Nr
      }

\maketitle

\section{Introduction}
The vibratory compaction of granular materials  has long been of
importance in technological applications, but it is only recently that
physicists have started to investigate the  process from a fundamental
perspective. In  particular, Knight et al.  performed
experiments \cite{KFLJN95} in  which a  column containing monodisperse
spherical beads  was subject  to  a long  sequence   of taps  with  an
intensity  characterized  by  $\Gamma=A/g$   where  $A$  is   the  maximum
acceleration and $g$ is the gravitational constant.  These experiments
showed that  the  density of the   beads increases  monotonically  and
surprisingly  slowly   with  the number   of  taps, $n_t$,  for  various
intensities of tapping. More specifically,  the density approaches the
steady state value as $1/\ln(n_t)$.

In  two    other  studies,  Nowak     et al.    \cite{NKPJN97,NKBJN98}
investigated the effect of cycling $\Gamma$, i.e. vibrating the column for
a fixed number of  taps with a  sequence of  $\Gamma$ values.  They  found
that as this parameter increases, the density of the granular material
increases. When $\Gamma$ is subsequently reduced  by reversing the initial
sequence, the density continues  to {\it increase}. This second branch
is reversible  in that it is   retraced if the  increasing sequence of
$\Gamma$ is repeated. In the same experiments, Nowak et al. monitored the
power spectrum of the density fluctuations near the steady state for
different values of $\Gamma$. The spectrum is distinctly non-Lorentzian,
with the highest and lowest characteristic frequencies being separated
by a non-trivial power-law-like regime. 

How can these effects be understood?   The energy required to displace
a  granular  particle through a  distance  equal  to its  diameter  is
typically  much  larger  than   $k_B  T$, so  granular  materials  are
essentially athermal systems  and  their properties are  dominated  by
geometrical frustration  effects.   When the system  is  already quite
dense,  even a small additional  increase in density  requires a large
scale  cooperative   rearrangement   of   the  particles.    The
logarithmic     compaction, the    irreversible/reversible  cycles  
and the non-Lorentzian power spectrum of the density fluctuations are
essentially the result of this effect.  

The experimental results have stimulated a few Monte Carlo studies
\cite{BM00}  and a plethora  of  simple  models that incorporate
geometrical frustration or quenched  disorder (see  Ref.\cite{H00} and
references    therein).    The   latter   include  the   (off-lattice)
adsorption-desorption or parking lot  model (PLM),  Frustrated Lattice
Gas (FLG) models, including Tetris, and one-dimensional lattice models
with short  range  dynamical  constraints\cite{PBS00}.  All these  are
capable of   reproducing qualitatively the   experimentally   observed
kinetics.

Different      Frustrated            Lattice             Gas     (FLG)
models\cite{CLHN97,NC99,N99,BL00,BL00b}  have been   proposed in which
particles occupy the sites of a square lattice tilted by 45$^o$.  Each
particle  has two internal degrees  of  freedom which,  in the  Tetris
realization\cite{CLHN97}, correspond to two possible orientations at a
site.  Neighboring sites can only be occupied if both particles have a
favorable  orientation.  Vibration    is introduced   by allowing  the
particles to move upward with  a probability $p_{up}$.  The connection
to  the  compaction  experiments   is    provided  by  the   parameter
$\gamma=-1/\ln(p_{up}/p_{down})$ which  plays the same role  as $\Gamma$.  The
main virtue  of  this description is that   it accounts for  the layer
structure of   the granular material.    In  the absence   of quenched
disorder, the system evolves to a compact  steady state according to a
$1/\ln(t)$-law.  To mimic  the  geometric frustration of  the packing,
kinetic constraints\cite{CLHN97,BL00}   can also be  replaced by
the introduction of quenched disorder\cite{NC99}.

Here we focus on the  adsorption-desorption  model in which hard  rods
are placed  on a line at randomly  selected positions  with a constant
rate  $k_{+}$.  If the trial  particle does not overlap any rod
already on the line, it is accepted.  In addition, all adsorbed rods are
subject to  removal   (desorption)  at  random  with  a  constant rate
$k_{-}$.  The properties  of the model  depend only on  the
ratio $K = k_{+}/k_{-}$,  with large values   of $K$ corresponding  to
small desorption rates.  (Pioneering work was  done in the limit
of $K\to\infty$\cite{JTT94,KB94}).   The connection to the  granular compaction
experiment is made by  regarding the particles on the  line as a layer
in the vibrating column.  The effect  of a tap  is to eject  particles
from the  layer (desorption), which  is followed by the replacement of
the particles,  generally  in   different positions,  and  possibly the
incorporation   of    additional particles.  $1/K$ plays a role
similar to that of $\Gamma$. In  this approach,  mechanical
stability is implicitly included by the absence of motion of the particles
when the desorption is switch off. This    microscopic and
off-lattice   description  of  granular  compaction    is suitable for
describing all previous experiments and, as we see below, is in fair
agreement  with  experimental results.

The kinetics  and equilibrium  properties of the adsorption-desorption
model            have         been        previously      investigated
\cite{NKBJN98,BKNJN98,KNT99,TTV99,TTV00}: in    the limit    of  small
desorption rate (that is appropriate to describe granular compaction),
the density increases  very slowly as $1/ \ln(t)$  until  it nears the
steady state  value, at  which point  the  kinetics cross over   to an
exponential form.  The   characteristic relaxation time $\tau_{eq}  $ of
this final   regime behaves   like $K^2/\ln(K)^3$ \cite{KNT99,TTV99}.
Moreover, the  density-density   correlation   function  exhibits  two
characteristic   time-scales:   the shortest  one   is proportional to
$\ln(K)$ and  is associated with a   local rearrangement of particles,
whereas  the  longest one,    $\tau_{eq}$, corresponds to   a collective
rearrangement  permitting an  increase of  the  density.  The ratio of
these time  scales  increases rapidly  as $K$ becomes  larger, and the
intermediate  timescale   region  is characterized    by a  power  law
behavior.

Here     we           show   that       the    irreversible/reversible
cycles\cite{NKPJN97,NKBJN98} and the recently demonstrated presence of
memory effects \cite{JTMJ00}  in the compaction  of glass beads can be
reproduced by the adsorption-model.   All the salient features of  the
phenomenology  of vibrated granular   materials are thus qualitatively
described by the model.  We also show that,  because of the very  slow
relaxation to equilibrium  (due  to cooperative rearrangements  of the
particles), the   model displays out-of-equilibrium  dynamical effects
such as aging\cite{BCKM98}, and we study  these properties both in the
response  and   the correlation functions.   Interestingly,  the aging
behavior of  the     two-time correlation  function  of    the density
fluctuations is found  to be different from  that observed for the FLG
and  Tetris   models\cite{NC99,BL00}. This  leads   us  to  suggest an
experimental way of  discriminating among the different models applied
to vibrated granular materials.

\section{Model and Simulation}
If time is expressed in units of $k_{+}$, the kinetics is given by
\begin{equation}\label{eq:1}
\frac{d\rho}{dt} = \Phi(t)-\frac{\rho}{K},
\label{kinet}
\end{equation}
where  $\Phi(t)$,  the insertion   probability at time  $t$  (or density
$\rho$),  is the  fraction of the  line  that is available for  the
insertion of a new particle.   The presence of a relaxation mechanism,
implies  that  the system   eventually   reaches a  steady state  that
corresponds  to an equilibrium   configuration of hard particles  with
$\rho_{\rm eq} = K\Phi_{\rm eq}(\rho_{\rm eq})$, where $\rho_{\rm eq}$ denotes
the equilibrium density.  At equilibrium, the insertion probability is
given exactly by
\begin{equation}\label{eq:2}
\Phi_{\rm eq}(\rho)= (1-\rho)\exp(-\rho/(1-\rho)).
\label{phieq}
\end{equation}
The kinetics    of the model  were   simulated  using the event-driven
algorithm  described previously  \cite{TTV00}.  

\section{Densification branches and memory effects}
\begin{figure}
\begin{center}

\resizebox{0.45\textwidth}{!}{\includegraphics{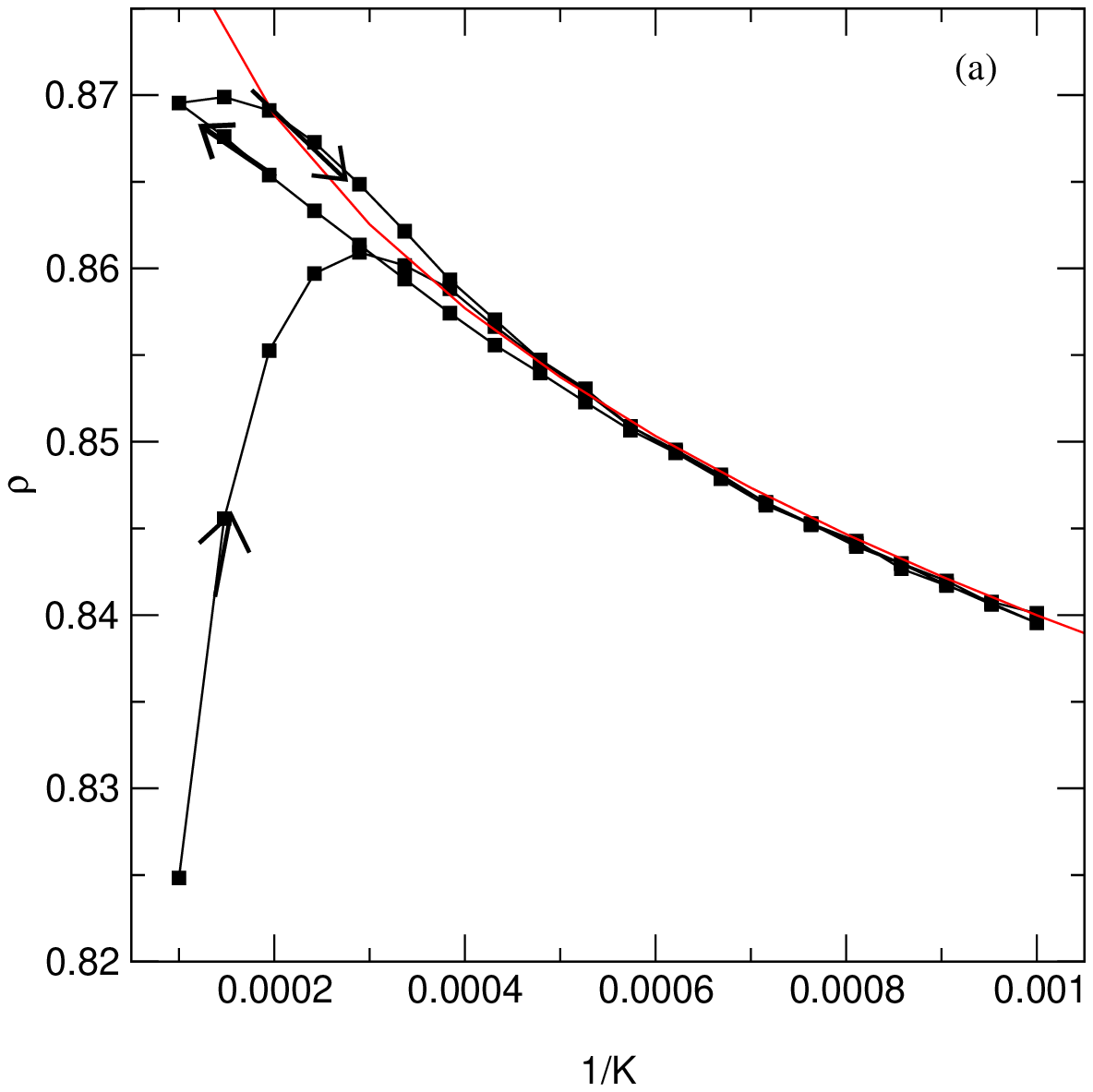}}
\resizebox{0.45\textwidth}{!}{\includegraphics{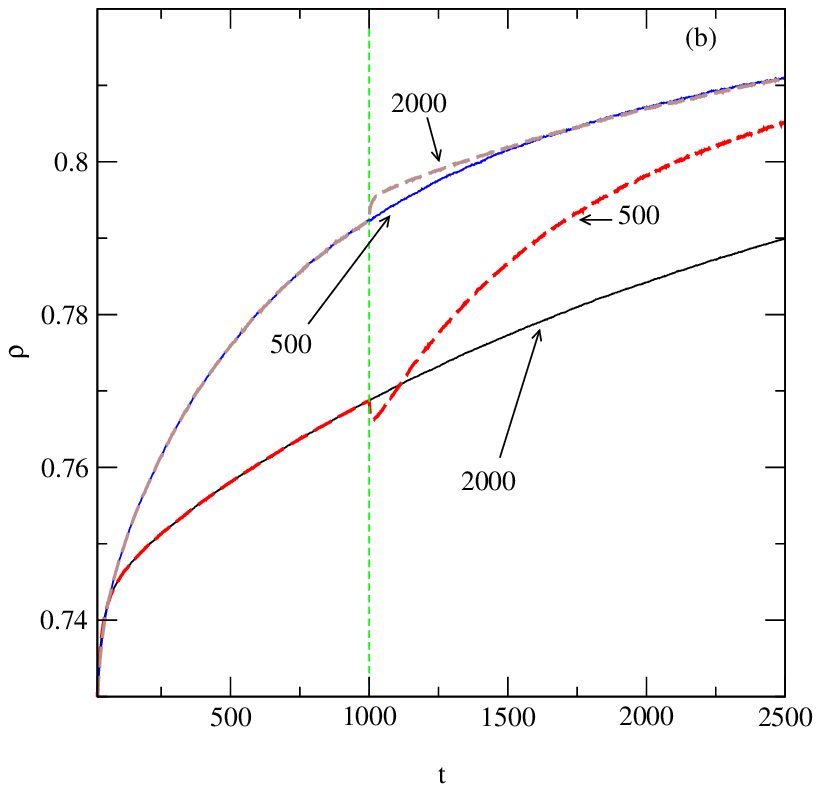}}

\caption{(a) Irreversible and reversible densification branches. Starting 
from a ``loose''  packed state, the  process consists of a sequence of
decreasing values of $K$.  For each $K$,  the duration is $40000$. The
arrows indicate the  way of cycle compaction.  The dotted  line is the
equilibrium  curve.  (b) Memory effect at   short times  after a rapid
change of the desorption rate: the full curves correspond to a process
with a constant $K$, whereas the dashed curves show  the kinetics of a
process where  $K$ is switched from $500$  to $2000$ and vice-versa at
$t=1000$}\label{fig:1}
\end{center}
\end{figure}

Following   the      experimental    procedure    of        compaction
cycles\cite{NKBJN98,NKPJN97}, we  have performed  simulations in which
the rate of desorption, $1/K$, is first increased  at a given rate and
then cycled down and up,  the simulation being  stopped after the same
time $t=40000$. (Recall that $1/K$ plays the  same role as the tapping
intensity and that, when expressed in units of  $k_+$, time is assumed
to measure the  number of taps; this  leads to the reasonable behavior
that the number of desorption events increases as the tapping strength
increases).  Figure\ref{fig:1}a  displays the density  as $1/K$ varies
between  $10^{-4}$  and  $10^{-3}$, and  is   similar to the  behavior
observed   experimentally   and      numerically     in     the    FLG
models\cite{NC99,BL00}.   Along  the  first (irreversible) branch  the
density  increases rapidly and then  passes through a maximum as $1/K$
increases.  When the  initial  sequence of  $1/K$  is  reversed ($1/K$
decreasing),  the density  increases monotonically.   When the initial
(increasing) sequence of $1/K$ is  repeated, the density now decreases
monotonically,  nearly  retracing  the  second  branch.   The residual
hysteresis  observed between the second   and third  branches is  also
present in other models\cite{H00}, and it  diminishes as one considers
larger  values of  $K$ and larger  time intervals   between changes of
desorption rate.   Note that  the  densities attained in   the present
adsorption-desorption model  are  typical of a  one-dimensional system
and  that more  realistic values  would be obtained   by employing the
two-dimensional version of the model.

To further compare the model predictions to the available experimental
data on vibrated granular materials, we  have studied the effect of an
abrupt change  in the  desorption  rate  $1/K$  on  the  densification
kinetics. This is illustrated in Fig. 1b in which $K$ is switched from
$500$  to $2000$ and vice-versa.  The same ``anomalous'' behavior as
observed  experimentally  by Josserand et   al. \cite{JTMJ00}, i.e. an
acceleration in the   densification  process when  $1/K$   is suddenly
lowered and the  reverse phenomenon when $1/K$  is increased, is found
in the adsorption-desorption model.

\section{Response functions and aging}

Compaction  cycles and memory effects characterize  the response of the
system to a moderate or large change of the tapping strength.  We consider
below another facet of the out-of-equilibrium dynamical behavior of
the system by examining the response to a small perturbation and the
associated two-time correlation function.

Non-equilibrium  studies  have  been     performed on  FLG  models:
Nicodemi\cite{N99} demonstrated  that these models  exhibit a negative
response and  Nicodemi and Coniglio \cite{NC99}  showed that the aging
regime of this model  is characterized by $C(t,t_w)\sim\ln(t_w)/\ln(t)$.
Barrat  and  Loreto\cite{BL00} performed extensive simulations  on the
Random Tetris model and clearly showed  that the aging behavior is the
same.

We calculate the response function as follows.  Starting from an empty
line, the system evolves at a fixed desorption rate.   At the end of a
fixed waiting period, $t_w$, two clones  of the system  are made.  The
original systems  continues to  evolve  with the same desorption  rate
$K$, and the two others with $K \pm \delta K$. The response then corresponds
to the difference in density of the copy  and the original system at a
later time  $t$.   In order to   obtain reasonable  statistics, it  is
necessary to average  over  many independent runs  (typically $10^4$).
Since the model   at equilibrium is equivalent   to a grand  canonical
ensemble,  the variable  conjugate   to the  density  is $\beta\mu$  where
$\beta=1/k_BT$ and $\mu$ is the chemical potential. From Eq.~(\ref{kinet})
at equilibrium we have that $\beta\mu=\ln(K)$ and the response function to
a change in $\ln(K)$ is therefore defined as
\begin{equation}\label{eq:3}
R(t,t_w)=\frac{\partial \rho(t)}{\partial \ln(K(t_w))}
\end{equation}
To be more precise, this is an integrated  response function since the
perturbation is applied over an extended  time.  Note that when $t$  becomes  larger than the  ergodic  time $\tau_{eq}$, the  response
function attains the  positive equilibrium  value $R_{eq}=\rho(1-\rho)^2$.
The fluctuation-dissipation  theorem applies when $t_w>t_{eq}$  and we
recover  the time   translation invariance $R(t,t_w)=R(\tau=t-t_w)$   in
which case  the   response function associated  with  changes  in  the
external field $\beta\mu=\ln(K)$ (coupled  to the density fluctuations)  and
the density-density correlation function are related:
\begin{equation}\label{eq:4}
R(\tau  )=\tilde{C}(0)-\tilde{C}(\tau )
\end{equation}
where $\tilde{C}(\tau )=<\delta\rho(\tau )\delta\rho(0)>$. (Note that as usual for
hard objects, the temperature is irrelevant and does not explicitly
enter in Eq.~(\ref{eq:4}) because it is included in $\ln(K)=\beta\mu$.)
From our previous study of the fluctuations around equilibrium
\cite{TTV00} we 
derive the analytical short time form which is exact for $K\to\infty$:
\begin{equation}\label{eq:5}
R(\tau )=\frac{L_W(K)^2}{1+L_W(K)}\left(1-\frac{L_W(K)}{\tau
+L_W(K)}e^{-\tau /K}\right)+
O(1/K)
\end{equation}
where  $L_W$ is   the  Lambert-W function   and  for  long  times  the
equilibrium value is attained exponentially with a characteristic time
of order  $K^2/   \ln(K)^3$.  It is   worth  stressing that with   the
definition   adopted above of  a  response  to  $\ln(K)$, a  positive
response means  that an increase of $K$  (i.e.,  a decrease in tapping
strength) increases   the  density,   whereas   a negative    response
corresponds to a density decrease.

\begin{figure}
\begin{center}
\resizebox{0.45\textwidth}{!}{\includegraphics{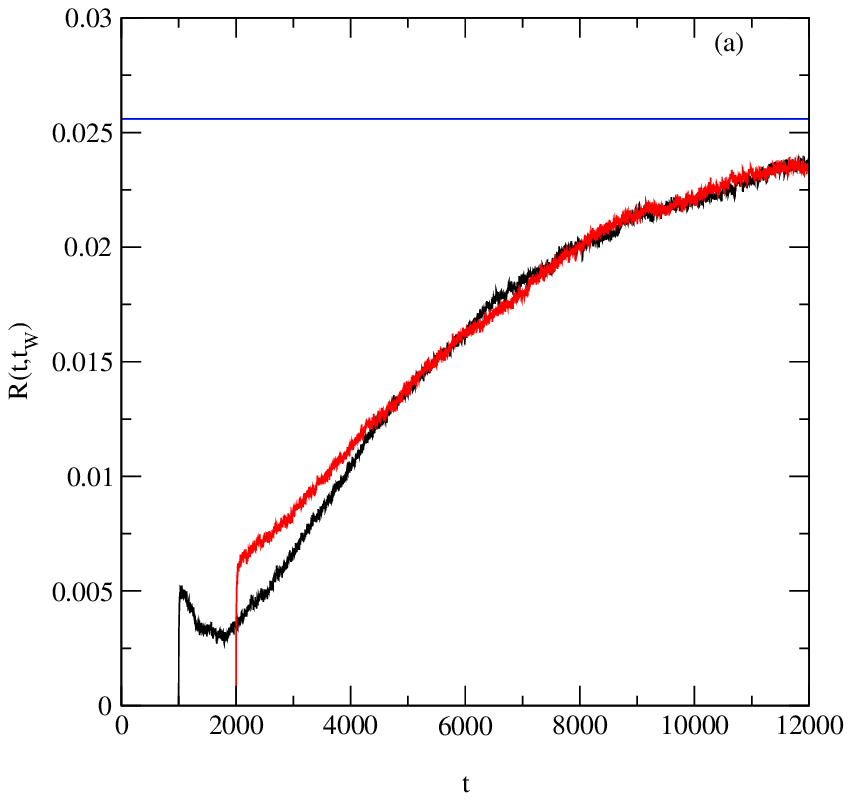}}
\resizebox{0.45\textwidth}{!}{ \includegraphics{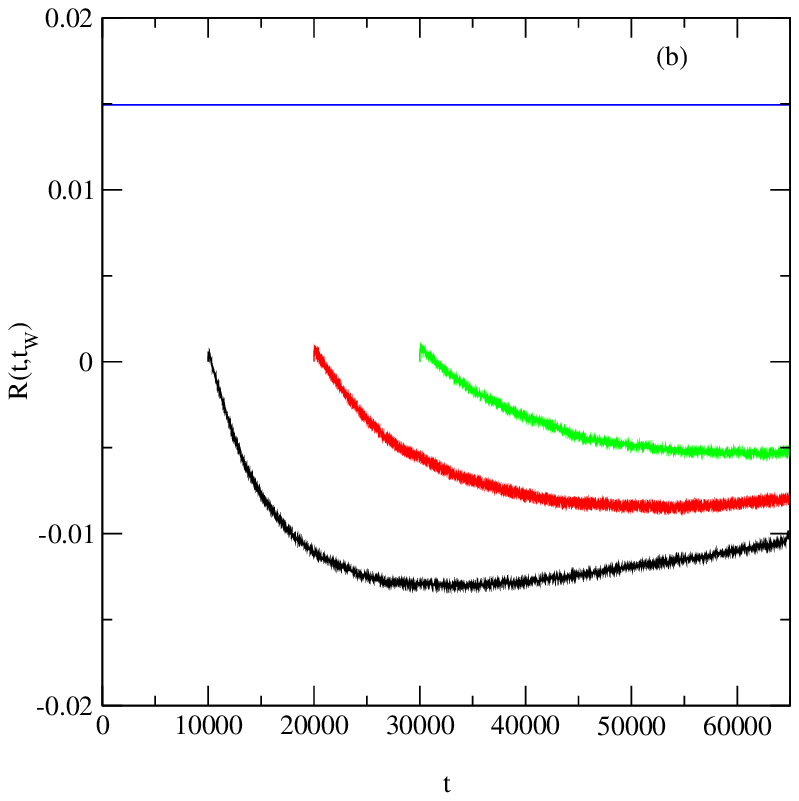}}

\caption{Two-time response function $R(t,t_w)$ as a function of
time $t$  for (a) $K=500$ with $t_w=1000, 2000$ and for (b) $K=5000$
with $t_w=10000, 20000, 30000$. The horizontal lines correspond to the
equilibrium value, $R_{eq}=\rho(1-\rho)^2$}\label{fig:2}
\end{center}
\end{figure}

Fig~\ref{fig:2}a displays  two  response functions for  a  system with
$K=500$.   For  a waiting time  ($t_w=2000$)  larger than  the ergodic
time\cite{BCKM98}, the system    is  already at equilibrium   and  the
response  is obviously monotonically increasing  and in agreement with
the  analytical  result,  Eq.   (\ref{eq:5}).    At  short times,  the
behavior  of both  curves is  quite  similar,  but for  $t_W=1000$ the
response then decreases and attains a  minimum before increasing again
towards   the  equilibrium value   (as    expected in  a  system  with
interrupted aging).  Fig~\ref{fig:2}b shows  the response functions in
the aging regime  for a larger value  of $K=5000$ for three  different
waiting    times.    The   shape  of     the  curves  is   similar  to
Fig~\ref{fig:2}a,  but the  amplitude    of the initial  increase   of
$R(t,t_w)$ is  smaller than for  $K=500$.  The negative  minimum, that
results from a  competition  between local rearrangements  (leading to
short-time  ``anomalous''    response  discussed above)    and  global
rearrangements (allowing for the densification), is less pronounced as
$t_w$ increases.   For long  times, which  are  not accessible in  the
simulations, the response   becomes   positive as it  approaches   the
equilibrium value.

In  order   to investigate  the   out-of-equilibrium evolution  of the
density-density            correlation  function,   we  consider   the
adsorption-desorption model  in the  limit   where the desorption   is
infinitely small\cite{JTT94}.  In  this case ($K\to\infty$), the relaxation
time $\tau_{eq}$ \cite{BCKM98} is infinite and the system always evolves
with out-of-equilibrium dynamics.  For $K$ large, but finite ($5000$),
the out of equilibrium and aging behavior is interrupted since $t$ may
become larger than $\tau_{eq}$.  We  have however checked that the aging
behavior when $t<\tau_{eq}$ is similar to that of the $K\to\infty$ case.

The   normalized  two-time  density-density   correlation function  is
defined as follows:
\begin{equation}
C(t,t_w) = \frac{<\rho(t)\rho(t_w)>-<\rho(t)><\rho(t_w)>}{<\rho(t_w)^2>-<\rho(t_w)>^2}
\end{equation}
where the angular brackets denote an average over independent runs and
$t\geq t_w$. 

\begin{figure}
\begin{center}
\resizebox{0.45\textwidth}{!}{\includegraphics{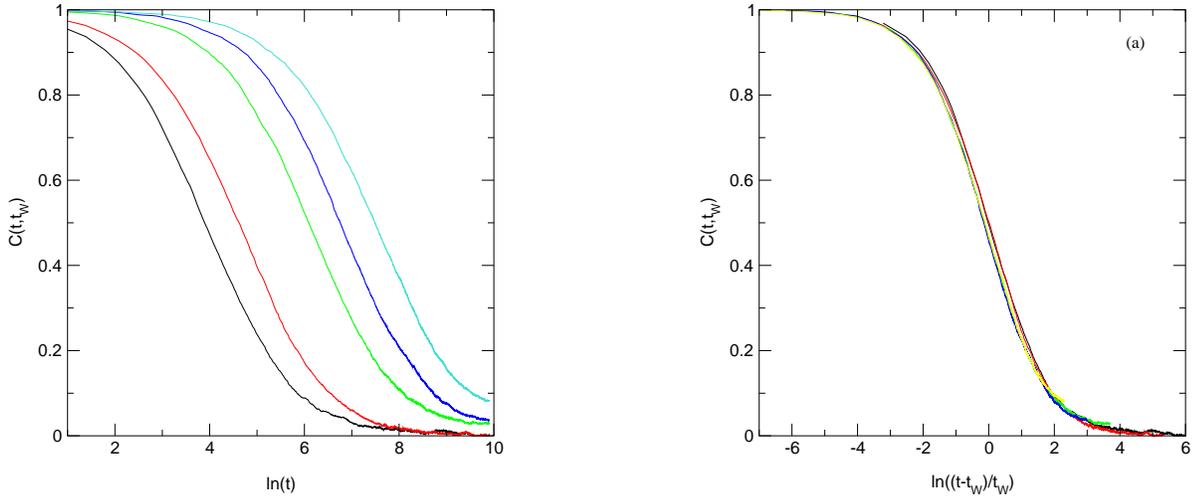}}

\caption{Two-time
density-density  correlator  $C(t,t_w)$  as a    function of time  for
various  waiting  times  $t_w$ ($t_w=50,100,500,1000,5000$   left   to
right). }\label{fig:3}
\end{center}
\end{figure}

Fig.~\ref{fig:3} shows $C(t,t_w)$  as a function  of $t$ for different
values  of  the waiting time $t_w$.    The system exhibits  aging: the
larger  $t_w$,  the longer the  memory   of the  initial configuration
persists. This  aging regime  is characterized  by  a violation of the
fluctuation-dissipation      theorem  and   the       breakdown     of
time-translational invariance \cite{BCKM98}.

\begin{figure}
\begin{center}
\resizebox{0.45\textwidth}{!}{\includegraphics{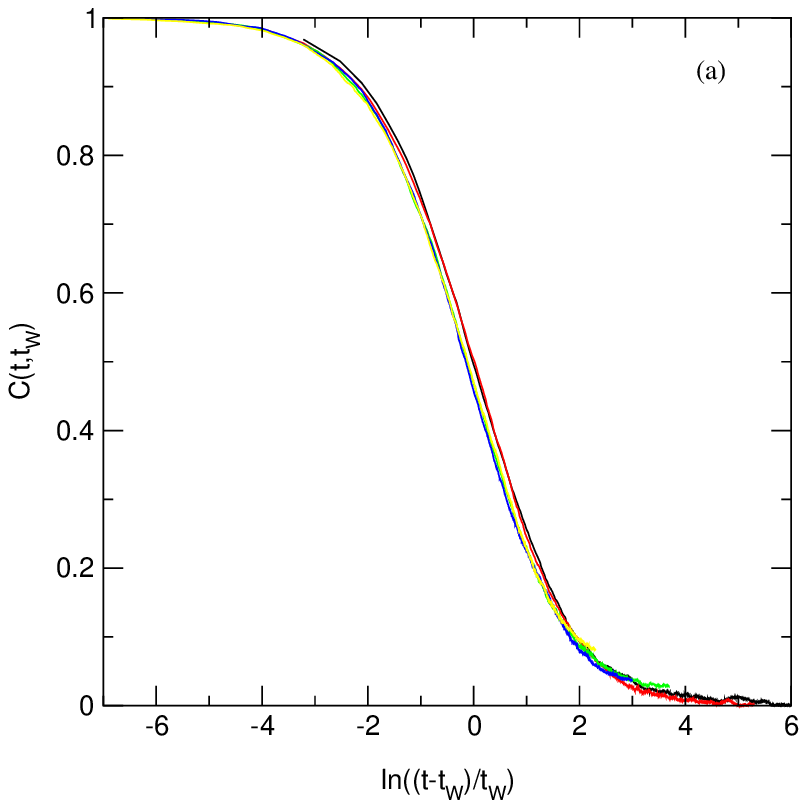}}
\resizebox{0.45\textwidth}{!}{ \includegraphics{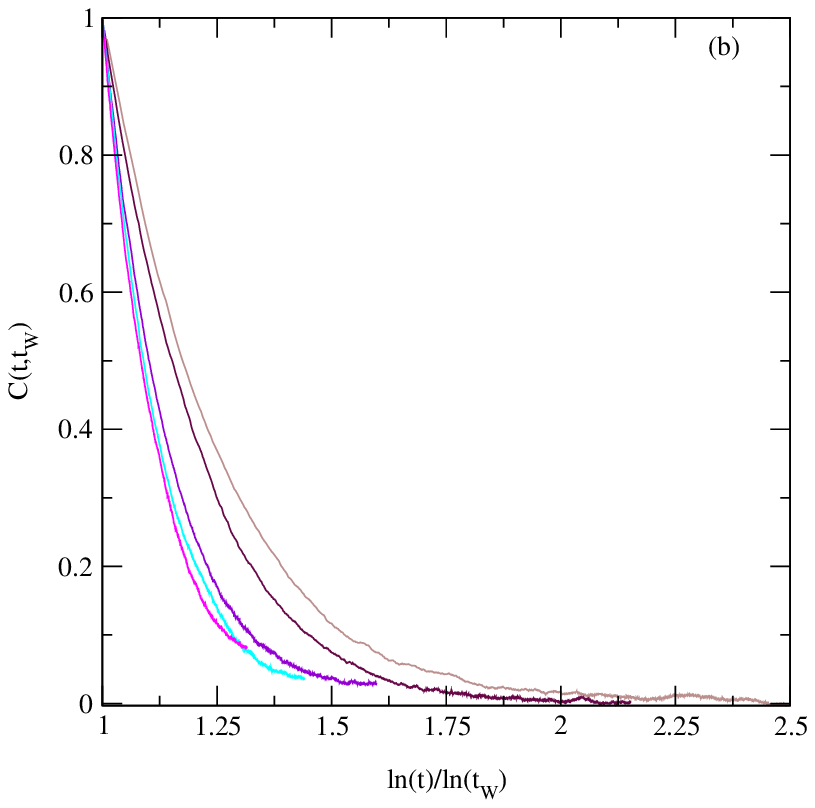}}
\caption{(a) Two-time
density-density correlator  $C(t,t_w)$ as a function  of (a)
$(t-t_w)/t_w$ and  (b)    $  \ln(t)/\ln(t_W)$  for  waiting  times
$t_w=50,100,500,1000,5000$). }\label{fig:4}
\end{center}
\end{figure}

When  $t$ and  $t_w$ are  large  enough  (but  still smaller than  the
equilibration time), the aging   behavior  is usually described by   a
non-trivial form involving a single scaling variable that is the ratio
of a function of $t$  (or of the  elapsed time $t-t_w$) divided by the
same function of  $t_w$. In the  FLG models, it  was found numerically
that        the      appropriate        scaling    variable         is
$\ln(t)/\ln(t_w)$\cite{NC99}. For  the  adsorption-desorption model we
show in Fig.~\ref{fig:4}a that when  the two-time correlation function
$C(t,t_w)$ is plotted as a function of $(t-t_w)/t_w$ the data for five
waiting times,  ranging from $25$   to $5000$, collapse onto  a single
curve.  This aging behavior, similar to that found in a large variety
of systems\cite{BCKM98}, is  thus different from the $\ln(t)/\ln(t_w)$
dependence of the  FLG models. Fig.~\ref{fig:4}b  clearly demonstrates
the absence of a master curve with this  scaling.  Moreover, no better
result  is obtained  when  $\ln(t-t_w)/\ln(t_w)$ is  used in place  of
$\ln(t)/\ln(t_w)$.

This    finding suggests  an  experimental   test  for deciding  which
phenomenological  model,   each   built   on   different  assumptions:
geometrical frustration in  the adsorption-desorption model,  quenched
disorder or kinetic constraints and layer  structure in FLG models, is
more relevant to describe the physics  of vibrated granular materials.
Since     all      predict  the     logarithmic       compaction,  the
irreversible/reversible cycles,  the non-Lorentzian power  spectrum of
the density  fluctuations      and  the  memory    effects    observed
experimentally, a promising  way to validate  or invalidate them as  a
minimal theoretical approach would be to study experimentally the
two-time correlation  function of the    density fluctuations in   the
course of the densification process.

\begin{acknowledgement}  
J.T.   thanks  the    National  Science  Foundation  (CHE-9814236) for
financial support.     P.V.     thanks A.  Barrat   for    stimulating
discussions.
\end{acknowledgement}

\end{document}